\documentclass[aps,prd,showkeys,preprint,nofootinbib,floatfix]{revtex4-1}
\pdfoutput=1

\usepackage{yfonts}
\usepackage{comment}
\usepackage{color}
\usepackage{amsmath}
\usepackage{amssymb}
\usepackage{amsthm}
\usepackage{mathrsfs}
\usepackage{graphicx}
\usepackage{fancyhdr}
\usepackage{array}
\usepackage{latexsym}
\usepackage[all]{xy}
\usepackage{eufrak}
\usepackage{euscript}
\usepackage{enumerate}
\usepackage{dsfont}
\usepackage{slashed}
\usepackage{hyperref}
\usepackage{caption}
\hypersetup{pdftex,colorlinks=true,linkcolor=blue,citecolor=blue,menucolor=black,urlcolor=blue,filecolor=blue}

\def\lsim{\mathop{\hbox{${\lower 3.8pt\hbox{$<$}}\atop{\raise 0.2pt\hbox{$\sim$}}$}}}
\def\gsim{\mathop{\hbox{${\lower 3.8pt\hbox{$>$}}\atop{\raise 0.2pt\hbox{$\sim$}}$}}}

\begin{document}

\title{Jet quenching parameters in strongly coupled anisotropic plasmas in the presence of magnetic fields}

\author{Romulo Rougemont}
\email{romulo.pereira@uerj.br, analisadorcetico at gmail dot com}
\affiliation{Departamento de F\'{i}sica Te\'{o}rica, Universidade do Estado do Rio de Janeiro,
Rua S\~{a}o Francisco Xavier 524, 20550-013, Maracan\~{a}, Rio de Janeiro, Rio de Janeiro, Brazil}

\begin{abstract}
I use the holographic gauge/gravity duality to systematically calculate the jet quenching parameters in strongly coupled anisotropic plasmas in the presence of external magnetic fields. The magnetic field breaks down spatial rotation symmetry from $SO(3)$ to $SO(2)$, leading to the presence of multiple anisotropic jet quenching parameters, which are evaluated here in two quite different holographic settings. One of them corresponds to a top-down deformation of the strongly coupled $\mathcal{N} = 4$ Super Yang-Mills plasma triggered by an external magnetic field, while the other one is a bottom-up Einstein-Maxwell-Dilaton model of phenomenological relevance for high energy peripheral heavy ion collisions, since it is able to provide a quantitative description of $(2+1)$-flavors lattice QCD thermodynamics with physical quark masses at zero and nonzero magnetic fields. I find for both models an overall enhancement of all the anisotropic jet quenching parameters with increasing magnetic fields. Moreover, I also conclude that for both models transverse momentum broadening is larger in transverse directions than in the direction of the magnetic field. Since these conclusions are shown to hold for two rather different holographic setups at finite temperature and magnetic fields, they are suggested as fairly robust features of strongly coupled anisotropic magnetized plasmas.
\end{abstract}

%\pacs{Valid PACS appear here}
% PACS, the Physics and Astronomy Classification Scheme.
% Valid PACS numbers may be entered using the \verb+\pacs{#1} command.

% Use showkeys class option if keyword display desired

\maketitle
\tableofcontents

%%%%%%%%%%%%%%%%%%%%%%%%%

\section{Introduction}
\label{intro}

The partons produced in high energy proton-proton (pp) and proton-nucleus (pA) collisions undergo multiple fragmentations, called parton showers, before hadronizing. The produced hadron jets typically contain many hadrons with high transverse momentum $p_T$ to the colliding beam axis. On the other hand, in high energy heavy ion collisions \cite{Arsene:2004fa,Adcox:2004mh,Back:2004je,Adams:2005dq,Aad:2013xma}, due to the formation of a deconfined medium dominated by color charges, called \emph{quark-gluon plasma} (QGP) \cite{Gyulassy:2004zy,Heinz:2013th,Shuryak:2014zxa}, the parton showers interact with the color charges of the medium changing the overall pattern observed for hadron jets relatively to the cases where no QGP is formed. Due to the energy loss of the partons traversing the QGP, there is a suppression of high $p_T$ jets in heavy ion collisions relatively to the cases of pp and pA collisions, what is called \emph{jet quenching} \cite{Gyulassy:1990ye,Wang:1991xy,Majumder:2010qh,Burke:2013yra}. For instance, imagine a $q\bar{q}$ pair traversing the QGP medium with some initial momentum $\vec{k}$. Due to the interaction with the other partons within the medium and the emission of gluon radiation, the two partons in the original $q\bar{q}$ pair may follow different paths inside the medium before reaching its boundaries, hadronizing into dijets. The parton which traveled the longer distance within the QGP looses more energy, therefore originating less high $p_T$ jets than the parton which traveled the shorter distance, leading to a dijet asymmetry characteristic of the jet quenching phenomenon. Indeed, jet quenching corresponds to one of the main experimental signatures of the QGP formation in heavy ion collisions \cite{Adler:2003ii,Adams:2003im,Back:2003ns,Arsene:2003yk}. During this process, each parton in the original $q\bar{q}$ pair has its initial momentum modified by the interaction with the medium, leading to an increase of their momentum $\vec{k}_\perp$ transverse to the initial momentum direction $\vec{k}$ of the pair,\footnote{Not to be confused with the transverse momentum to the colliding beam axis, $p_T$.} what is called \emph{transverse momentum broadening}. This is associated with the radiative energy loss of highly energetic partons (hard probes) traversing the QGP medium and can be characterized by the so-called \emph{jet quenching parameter}, $\hat{q}$ \cite{Baier:1996sk,Zakharov:1997uu}, which is perturbatively defined as the transverse momentum diffusion constant corresponding to the fraction of mean transverse momentum squared gained by the hard probe within the medium per unit length trajectory, $\hat{q}=d\langle k_\perp^2\rangle/dL$, where $L$ is the distance traveled by the parton within the QGP.

The QGP produced in relativistic heavy ion collisions at RHIC and the LHC at temperatures not far above the QCD crossover transition \cite{Aoki:2006we,Borsanyi:2016ksw} is known to be a strongly coupled medium, since phenomenological relativistic viscous hydrodynamical models can quantitatively describe a wealth of experimental heavy ion data by using very small values of shear viscosity \cite{Heinz:2013th,Ryu:2015vwa,Bernhard:2016tnd,Bernhard:2018hnz}, which is seen as an universal feature of strongly coupled media \cite{Policastro:2001yc,Buchel:2003tz,Kovtun:2004de}. Therefore, it is a task of phenomenological interest to calculate the jet quenching parameter in strongly coupled settings.

In this regard, the holographic gauge/gravity duality \cite{Maldacena:1997re,Gubser:1998bc,Witten:1998qj,Witten:1998zw} constitutes a nonperturbative framework to map physical observables of some strongly coupled quantum field theories into calculations involving classical general relativity in higher dimensional asymptotically AdS spacetimes (see Refs. \cite{CasalderreySolana:2011us,Adams:2012th} for some recent reviews with many phenomenological applications). Particularly, in the case of light partons travelling through a strongly coupled medium, a \emph{nonpeturbative definition} of the jet quenching parameter $\hat{q}$ has been originally proposed in Refs. \cite{Liu:2006ug,Liu:2006he} (see also Ref. \cite{DEramo:2010wup} for updated discussions), based on the calculation of light-like adjoint Wilson loops, and then applied to calculate the jet quenching parameter in the so-called $\mathcal{N}=4$ Super Yang-Mills (SYM) plasma at finite temperature. Since then, this prescription has been employed to calculate the jet quenching parameter in holographic models with finite 't Hooft coupling corrections \cite{Armesto:2006zv,Zhang:2012jd}, anisotropic sources \cite{Giataganas:2012zy,Chernicoff:2012gu,Ammon:2012qs,Li:2016bbh}, including higher order derivative corrections of the bulk metric \cite{Misobuchi:2015ioa}, and nonconformal settings \cite{DeWolfe:2009vs,Gursoy:2009kk,Li:2014hja,Rougemont:2015wca}.\footnote{This is by no means an exhaustive list of references on the topic.} It is also important to mention that a different definition of the jet quenching parameter for heavy quarks, related to the Langevin dynamics and momentum fluctuations of heavy probes, has been also investigated in holographic settings, see e.g. Refs. \cite{Herzog:2006gh,Gubser:2006nz}.

For possible realistic phenomenological applications of the gauge/gravity duality to QCD environments, it is essential that conformal symmetry is dynamically broken in the holographic setup, since the dynamical generation of the $\Lambda_{\textrm{QCD}}$ scale is a distinctive feature of QCD. Indeed, a simple comparison between the conformal thermodynamics and hydrodynamics of the SYM plasma with their highly nonconformal counterparts found in the QGP not far above the crossover region (which is the relevant region for heavy ion phenomenology) shows that the SYM plasma is radically different from the real-world QGP produced in heavy ion collisions \cite{Rougemont:2016etk}. This fact was the main motivation that lead originally to the formulation of dynamical dilatonic holographic models emulating the behavior of the actual QGP \cite{Gubser:2008ny,Gubser:2008yx,Gubser:2008sz,DeWolfe:2010he,DeWolfe:2011ts} (for earlier developments concerning the vacuum, see Refs. \cite{Csaki:2006ji,Gursoy:2007cb,Gursoy:2007er}). The dilaton field in such approaches is responsible for dynamically breaking the conformal symmetry in the holographic setup, with the dilaton potential being engineered in order to constrain the background black hole solutions to reproduce some phenomenological inputs. The main purpose of such endeavor is not merely reproduce via holography actual data of real-world physical systems, but mainly provide new \emph{predictions} for observables which could be tested by comparison with either first principle calculations (e.g., lattice QCD simulations) or experimental data (which requires using the microscopic outputs generated by such holographic models in phenomenological codes used to describe, for instance, the spacetime evolution of the medium produced in heavy ion collisions).

By further developing the main ideas of these early works, in Ref. \cite{Finazzo:2014cna} there was proposed an Einstein-Dilaton model constructed to match lattice QCD equation of state at zero chemical potential and vanishing electromagnetic fields. The model was employed in this reference to predict the temperature dependence of several transport coefficients of second order nonconformal relativistic viscous hydrodynamics \cite{Romatschke:2009kr}. Besides the smallness of the shear viscosity to entropy density ratio, $\eta/s=1/4\pi$, which is naturally enclosed in holographic models and reflects the strongly coupled nature of the QGP not far above the crossover region, a remarkable \emph{prediction} of Ref. \cite{Finazzo:2014cna} was the temperature dependence of the bulk viscosity to entropy density ratio, $\zeta/s$, which matches fairly well recent profiles for this observable favored in Bayesian analysis of hydrodynamic models simultaneously describing several heavy ion data \cite{Bernhard:2016tnd,Bernhard:2018hnz}. This model was extended in Refs. \cite{Rougemont:2015wca,Rougemont:2015ona,Finazzo:2015xwa,Rougemont:2017tlu,Critelli:2017oub,Rougemont:2018ivt} to an Einstein-Maxwell-Dilaton (EMD) model, which was then used to predict the behavior of several physical observables as functions of temperature $T$ and baryon chemical potential $\mu_B$. In these references, there were obtained quantitative agreement of the EMD \emph{predictions} for the finite temperature and baryon density equation of state and the higher order baryon susceptibilities with the corresponding state-of-the-art first principles lattice QCD results \cite{Bazavov:2017dus,Borsanyi:2018grb}. These results illustrate some of the actual capabilities of EMD holography (with bulk actions adequately constrained by some phenomenological input data) in what regards applications to real-world physical systems, like the QGP produced in heavy ion collisions.

Of central importance to the present work, there is the fact that in high energy peripheral heavy ion collisions there are attained the highest values of magnetic fields ever produced by the humankind, whose estimates for the earliest stages of ultrarelativistic noncentral collisions typically range from $eB\sim m_\pi^2\sim 0.02 \, \mathrm{GeV^2}$ at RHIC to $eB\sim 15 m_\pi^2\sim 0.3 \, \mathrm{GeV^2}$ at the LHC \cite{Skokov:2009qp,Deng:2012pc,Bloczynski:2012en,Tuchin:2013ie,Bali:2011qj}.\footnote{One expects that these strong magnetic fields have significantly decayed at the time the QGP is formed (roughly $\sim 1$ fm/c after the collision) because of the receding spectators leaving the collision zone. However, one also needs to take into account that the electric conductivity \cite{Tuchin:2013apa,Gursoy:2014aka} and the quantum nature of the sources \cite{Holliday:2016lbx} may significantly delay this decay within the medium. Therefore, it is not clear at present the extent to which the large magnetic fields generated at the earliest stages of noncentral heavy ion collisions can actually affect the hydrodynamics and the thermodynamics of the QGP formed at later stages.} In this case, even the medium taken in thermodynamic equilibrium is no longer isotropic because spatial rotation symmetry is broken down from $SO(3)$ to $SO(2)$ in the plane orthogonal to the direction of the external magnetic field, and one has to consider the calculation of multiple jet quenching parameters.

In this work, I focus on the investigation of the jet quenching parameters in strongly coupled anisotropic fluids in the presence of external magnetic fields, as described by two different holographic models.

\newpage

The first model corresponds to a top-down deformation of the SYM plasma driven by an external magnetic field, called the ``magnetic brane model'' \cite{DHoker:2009mmn}. Although this model is nonconformal, the breaking of the conformal symmetry is explicitly done by the magnetic field \cite{Fuini:2015hba}, and therefore there is no analogous of the QCD dynamical symmetry breaking in this case (once the magnetic field is switched off, the model becomes conformal again). In Ref. \cite{Li:2016bbh} the anisotropic jet quenching parameters were calculated for an analytical approximation of the magnetic brane background strictly valid for the limit $eB/T^2\gg 1$. Here I go beyond this analytical limit and evaluate the complete results for the anisotropic jet quenching parameters of the magnetic brane model valid for any value of $eB/T^2$ by making use of the full numerical solutions for this background, which correspond to a holographic renormalization group flow from a BTZ$\,\otimes\,\mathbb{R}^2$ \cite{Banados:1992wn} black hole in the infrared to the AdS$_5$ geometry in the ultraviolet (for other calculations involving this model, see for instance Refs. \cite{Fuini:2015hba,Basar:2012gh,Critelli:2014kra,Rougemont:2014efa,Finazzo:2016mhm}).

The main results of the present work regard the calculation of the anisotropic jet quenching parameters in the phenomenological holographic EMD model with magnetic fields originally proposed in Refs. \cite{Finazzo:2016mhm,Rougemont:2015oea,Critelli:2016cvq}. In these works the magnetic EMD model has been shown to be able to correctly \emph{predict} the quantitative behavior of the lattice QCD equation of state at finite temperature and magnetic field \cite{Bali:2014kia}, alongside with the entropy of a heavy quark \cite{Bazavov:2016uvm} and the renormalized Polyakov loop \cite{Bruckmann:2013oba,Endrodi:2015oba} in the deconfined QGP phase, reinforcing the reach of capabilities of phenomenological EMD holography.

This manuscript is organized as follows. In section \ref{secANISO} I review the general holographic formulas for the anisotropic jet quenching parameters. In section \ref{secDK} I review the basics of the magnetic brane model and present the full results for the jet quenching parameters in this background, valid for any value of $eB/T^2$. In section \ref{secEMD} I review the basics of the phenomenological magnetic EMD model and evaluate the corresponding anisotropic jet quenching parameters as functions of temperature and magnetic field. It will be shown that in both models there is an overall enhancement of all the anisotropic jet quenching parameters with increasing magnetic fields and that transverse momentum broadening of light partons is higher in transverse directions than in the direction of the magnetic field.

I use in this work natural units with $c=\hbar=k_B=1$ and a mostly plus (Lorentzian) metric signature.

\section{General holographic formulas for the anisotropic jet quenching parameters}
\label{secANISO}

The jet quenching parameter for light partons can be calculated through light-like adjoint Wilson loops, which for isotropic holographic fluids was first considered in Refs. \cite{Liu:2006ug,Liu:2006he}. The generalization of this approach for anisotropic holographic media was originally pursued in Refs. \cite{Giataganas:2012zy,Chernicoff:2012gu}. In this section I summarize the main formula obtained in Ref. \cite{Giataganas:2012zy} for three relevant configurations of the jet quenching parameter in anisotropic settings, namely, when the light parton is moving parallel to the direction of the anisotropy source (in the present work, an external magnetic field), and when the light parton is moving perpendicular to the magnetic field, in which case one may consider the transverse momentum broadening taking place in the same direction of the magnetic field, or in a direction perpendicular to the magnetic field.\footnote{Regarding the general formula for a light parton moving in an arbitrary direction relatively to the anisotropy source, see Ref. \cite{Chernicoff:2012gu}.}

Below, I promptly adapt this general formula for a generic radial coordinate $\tilde{r}$ where the boundary of the bulk geometry lies at infinity (as in the case of the models considered in the present manuscript) and already write it in terms of the background functions expressed in the Einstein frame, making explicit the contribution of the background dilaton field $\tilde{\phi}(\tilde{r})$,
\begin{align}
\frac{\hat{q}_{p(k)}}{\sqrt{\lambda_t}T^3} = \frac{1}{\pi T^3} \left( \int_{\tilde{r}_H}^{\tilde{r}_\textrm{max}} d\tilde{r}\, \frac{1}{\tilde{g}^{(s)}_{kk}}\, \sqrt{\frac{\tilde{g}^{(s)}_{rr}}{\tilde{g}^{(s)}_{tt}+\tilde{g}^{(s)}_{pp}}} \right)^{-1}
 = \frac{1}{\pi T^3} \left( \int_{\tilde{r}_H}^{\tilde{r}_\textrm{max}} d\tilde{r}\, \frac{1}{e^{\sqrt{2/3}\,\tilde{\phi}}\,\tilde{g}_{kk}}\, \sqrt{\frac{\tilde{g}_{rr}}{\tilde{g}_{tt}+\tilde{g}_{pp}}} \right)^{-1},
\label{eq:qhatgen}
\end{align}
where $\lambda_t$ is the 't Hooft coupling of the boundary quantum gauge theory, $p$ denotes the direction of movement of the light parton within the strongly coupled medium, $k$ is the direction considered for the transverse momentum broadening (which, by definition, is always perpendicular to the direction $p$), the tilde regards the background functions written in the so-called ``standard coordinates'' (to be discussed in sections \ref{secDK} and \ref{secEMD}), $\tilde{r}_H$ is the radial location of the background black hole horizon, $\tilde{r}_\textrm{max}$ is the radial location of the boundary (which formally goes to infinity), and the relation between the metric expressed in the string and Einstein frames for the normalization of the EMD action to be discussed in section \ref{secEMD} is given by \cite{Rougemont:2015wca},\footnote{If one considers instead the normalization for the dilaton $\tilde{\varphi}$ used, for instance, in Ref. \cite{Gursoy:2009kk}, this relation would read $\tilde{g}_{\mu\nu}^{(s)}=e^{4\tilde{\varphi}/3}\,\tilde{g}_{\mu\nu}$, where by comparison $\tilde{\varphi}=\sqrt{3/8}\,\tilde{\phi}$ \cite{Rougemont:2015wca}.}
\begin{align}
\tilde{g}_{\mu\nu}^{(s)}=e^{\sqrt{2/3}\,\tilde{\phi}}\,\tilde{g}_{\mu\nu}.
\label{eq:stringeinstein}
\end{align}

I am going to consider in this work the external magnetic field in an arbitrary $z$ direction. The background metric has $SO(2)$ rotation symmetry in the transverse plane, therefore, $\tilde{g}_{xx}=\tilde{g}_{yy}$.

\subsection{Light parton moving parallel to the magnetic field}
\label{subsecPar}

For a light parton moving in the same direction $z$ of the magnetic field ($p=z$) and transverse momentum broadening taking place in the transverse plane to the movement ($k=x,y$), one has from Eq. \eqref{eq:qhatgen},
\begin{align}
\frac{\hat{q}_{\parallel(\perp)}}{\sqrt{\lambda_t}T^3} = \frac{1}{\pi T^3} \left( \int_{\tilde{r}_H}^{\tilde{r}_\textrm{max}} d\tilde{r}\, \frac{1}{e^{\sqrt{2/3}\,\tilde{\phi}}\,\tilde{g}_{xx}}\, \sqrt{\frac{\tilde{g}_{rr}}{\tilde{g}_{tt}+\tilde{g}_{zz}}} \right)^{-1}.
\label{eq:qParPerp}
\end{align}

\subsection{Light parton moving perpendicular to the magnetic field}
\label{subsecPerp}

For a light parton moving perpendicular to the magnetic field, one needs to take into account an important observation pointed out in Ref. \cite{Li:2016bbh}. Due to the definition of the jet quenching parameter in the isotropic case, $\hat{q}_{\textrm{(iso)}}$, which regards the momentum diffusion in the transverse plane to the direction of movement of the light parton, and due to the fact that one must recover $\hat{q}_{\textrm{(iso)}}$ when the magnetic field is switched off, one must satisfy the following constraint,
\begin{align}
\hat{q}_{\textrm{(iso)}} = \lim_{B\to 0}\hat{q}_{\parallel(\perp)}=\lim_{B\to 0} \left[\hat{q}_{\perp(\parallel)} + \hat{q}_{\perp(\perp)}\right] = 2 \lim_{B\to 0} \hat{q}_{\perp(\parallel)},
\label{eq:constraint}
\end{align}
where $\hat{q}_{\perp(\parallel)}$ [$\hat{q}_{\perp(\perp)}$] denotes the contribution to the jet quenching parameter of the light parton moving perpendicular to the magnetic field (e.g., $p=x$) coming from the transverse momentum broadening taking place in the direction of the magnetic field ($k=z$) [in the other direction perpendicular to the magnetic field ($k=y$)]. That is, due to the isotropy symmetry breaking promoted by the magnetic field, in face of the constraint \eqref{eq:constraint} one needs to consider an extra factor of $1/2$ in front of Eq. \eqref{eq:qhatgen} when calculating if for a light parton moving perpendicular to the anisotropy source (magnetic field).\footnote{Notice that Eq. \eqref{eq:qParPerp} already computes the total contribution from transverse momentum broadening in the plane perperdicular to the light parton moving in the direction of the magnetic field, since in this case there is a $SO(2)$ symmetry in this plane. If one wishes to compute the separate momentum diffusions in the $x$ or $y$ direction, one just needs to multiply the result by $1/2$.} Therefore, one has,
\begin{align}
\frac{\hat{q}_{\perp(\parallel)}}{\sqrt{\lambda_t}T^3} &= \frac{1}{2\pi T^3} \left( \int_{\tilde{r}_H}^{\tilde{r}_\textrm{max}} d\tilde{r}\, \frac{1}{e^{\sqrt{2/3}\,\tilde{\phi}}\,\tilde{g}_{zz}}\, \sqrt{\frac{\tilde{g}_{rr}}{\tilde{g}_{tt}+\tilde{g}_{xx}}} \right)^{-1},\label{eq:qPerpPar}\\
\frac{\hat{q}_{\perp(\perp)}}{\sqrt{\lambda_t}T^3} &= \frac{1}{2\pi T^3} \left( \int_{\tilde{r}_H}^{\tilde{r}_\textrm{max}} d\tilde{r}\, \frac{1}{e^{\sqrt{2/3}\,\tilde{\phi}}\,\tilde{g}_{xx}}\, \sqrt{\frac{\tilde{g}_{rr}}{\tilde{g}_{tt}+\tilde{g}_{xx}}} \right)^{-1},\label{eq:qPerpPerp}
\end{align}
where in the last equation I used the fact that for the backgrounds considered here, $\tilde{g}_{yy}=\tilde{g}_{xx}$.

In the isotropic limit of zero magnetic field, one has $\tilde{g}_{xx}=\tilde{g}_{zz}$, and it becomes clear from Eqs. \eqref{eq:qParPerp}, \eqref{eq:qPerpPar}, and \eqref{eq:qPerpPerp} that the constraint \eqref{eq:constraint} is satisfied.

\section{The magnetic brane model}
\label{secDK}

\subsection{The holographic model}
\label{secDKmodel}

In this section I briefly review the basics of the magnetic brane model \cite{DHoker:2009mmn}, regarding what is just necessary for the computation of the corresponding numerical backgrounds and the anisotropic jet quenching parameters.

The background dilaton field is zero in this model, $\tilde{\phi}=0$, and the bulk Einstein-Maxwell action is given by,
\begin{equation}
\label{eq:magbranesaction}
S = \frac{1}{16\pi G_5}\int_{\mathcal{M}_5}d^5x\sqrt{-g}\left[R + \frac{12}{l_\textrm{AdS}^2}- F_{\mu\nu}^2\right],
\end{equation}
where $l_\textrm{AdS}$ is the asymptotic AdS$_5$ radius, which I set to unity from now on. This action is supplemented by boundary terms related to the holographic renormalization of the model (which will not be needed in the calculations pursued here, therefore I omit them), plus a topological 5D Chern-Simons term which vanishes on-shell for the background solutions to be discussed next. The Chern-Simons term, however, does play a role in this model in what regards the identification of the relation between the bulk magnetic field (denoted in the present section by $B$) and the physically observable boundary magnetic field (denoted in the present section by $\mathcal{B}$). This relation reads $\mathcal{B} = \sqrt{3} B$ \cite{DHoker:2009mmn}.

In the so-called ``standard coordinates'' (denoted by a tilde), the ansatz for the anisotropic magnetic brane background with an external constant magnetic field $\vec{B}=B\hat{z}$ is given by,
\begin{equation}
\label{eq:magbranatz}
ds^2=-\tilde{U}(\tilde{r})d\tilde{t}^2+\frac{d\tilde{r}^2}{\tilde{U}(\tilde{r})} + e^{2\tilde{V}(\tilde{r})}(d\tilde{x}^2 + d\tilde{y}^2) + e^{2\tilde{W}(\tilde{r})} d\tilde{z}^2, \quad F = B d\tilde{x} \wedge d\tilde{y}.
\end{equation}
Maxwell's equations are trivially satisfied by this ansatz and Einstein's equations can be worked out to give the following set of coupled ordinary differential equations,
\begin{align}
U''(r)+\frac{5}{3} U'(r) \left(2 V'(r)+W'(r)\right)+\frac{4}{3} U(r) V'(r) \left(V'(r)+2 W'(r)\right)-16 &= 0, \\
V''(r)+\frac{V'(r) \left(-U'(r)-5 U(r) W'(r)\right)}{3 U(r)}+\frac{12-2 U'(r) W'(r)}{3 U(r)}+\frac{2}{3} V'(r)^2 &= 0, \\
W''(r)+\frac{W'(r) \left(4 U'(r)+10 U(r) V'(r)\right)}{3 U(r)}+\frac{2 U'(r) V'(r)+2 U(r) V'(r)^2-24}{3 U(r)}+W'(r)^2 &= 0.
\end{align}

In order to numerically solve the above equations of motion, one introduces first a new set of coordinates, the so-called ``numerical coordinates'' (denoted without tildes), which are needed to ascribe numerical values to all the initial data at the black hole horizon required to start the numerical integration of the set of coupled differential equations. This can be done as follows: one fixes the radial location of the black hole horizon at unity in the numerical radial coordinate, $r_H=1$, which implies $U(1)=0$; one may also choose the numerical time coordinate such that $U'(1)=1$, which implies that Hawking's temperature is given by,
\begin{align}
T=\frac{\sqrt{-\tilde{g}'_{tt}\tilde{g}^{rr}\,'}}{4\pi}\biggr|_{\tilde{r}=\tilde{r}_H}= \frac{\sqrt{\tilde{U}'(\tilde{r})^2}}{4\pi}\biggr|_{\tilde{r}=\tilde{r}_H} = \frac{|U'(1)|}{4\pi} = \frac{1}{4\pi}.
\label{eq:DKtemp}
\end{align}
By rescaling the $(\tilde{x},\tilde{y})\mapsto(x,y)$ coordinates one may fix $V(1)=0$, while rescaling the $\tilde{z}\mapsto z$ coordinate one may fix $W(1)=0$. In these rescaled coordinates the magnetic field is denoted by $b$, which is taken as an initial condition, such that for each chosen value of $b$ it is generated a numerical solution corresponding to some specific physical state at the dual boundary quantum field theory. By Taylor expanding the background functions near the horizon one can show that at lowest order \cite{DHoker:2009mmn},
\begin{align}
V'(1) = 4-\frac{4}{3} b^2 \quad \mathrm{and} \quad W'(1) = 4+\frac{2}{3} b^2.
\end{align}
With the horizon data specified as above, one can now numerically integrate the equations of motion for different values of the initial condition $b$. In order to avoid the singular point of the equations of motion at the horizon in the numerical routine, one may start the numerical integration slightly beyond the horizon, at $r_{\textrm{start}}=1+\epsilon$, with $\epsilon=10^{-5}$, for instance. Formally, in these coordinates the boundary is at infinity. However, in practice, for numerical integration it is impossible to go to infinity, so one must stop it at some ``large'' value of $r$. The criterion which decides how much large the value of $r$ needs to be in order to stop the numerical integration is the behavior of the numerical background, which must asymptote to AdS$_5$ in the ultraviolet. This can be checked by evaluating the Ricci scalar on top of the numerically generated backgrounds. For $b>0$ and small $r$, the Ricci scalar will generally be different from its ultraviolet AdS$_5$ value, $R(r\to\infty)=-20$. For small values of $b$, the Ricci scalar reaches its AdS$_5$ value already for values of the radial coordinate $r$ of order $\lesssim 10$. As one increases the value of the initial condition $b$, the ultraviolet fixed point corresponding to the AdS$_5$ geometry is attained at larger values of $r$. For $b\ge\sqrt{3}$ it is no longer possible to find asymptotically AdS$_5$ geometries \cite{DHoker:2009mmn},\footnote{At $b=\sqrt{3}$ the numerical background gives a constant Ricci scalar equals to $-18$ for any value of $r$, so this does not correspond to an asymptotically AdS$_5$ geometry.} therefore the range of the rescaled bulk magnetic field must be $b\in[0,\sqrt{3})$. Up to $b\sim 1.371$ the ultraviolet fixed point is reached around $r\sim 10^2$. Continuing the numerical integration of the equations of motion for larger values of $r$ beyond the point where the ultraviolet fixed point is reached does not change the background geometry anymore and constitutes a waste of time. So one can stop it, for instance, at $r_{\textrm{max}}=10^5$ (if one wishes to use values of $b$ very close to the critical value $\sqrt{3}$).

One can check that the numerical background functions asymptote to $U(r\to\infty,b)\sim r^2$, $e^{2V(r\to\infty,b)}\sim v(b)r^2$, $e^{2W(r\to\infty,b)}\sim w(b) r^2$, where one can numerically obtain $v(b)\equiv e^{2V(r_{\textrm{max}},b)}/r_{\textrm{max}}^2$ and $w(b)\equiv e^{2W(r_{\textrm{max}},b)}/r_{\textrm{max}}^2$. The numerical factors $v(b)$ and $w(b)$ deviate the form of the AdS$_5$ metric written in the numerical coordinates from its standard form, in terms of which standard holographic formulas are derived. To use these standard formulas, one must then rescale the spatial numerical coordinates back to the standard ones, so that the relations between the standard and the numerical coordinates are given as follows \cite{DHoker:2009mmn},
\begin{align}
\tilde{t} &= t, \,\,\, \tilde{r} = r, \,\,\,\tilde{x} = \frac{x}{\sqrt{v(b)}}, \,\,\,\tilde{y} = \frac{y}{\sqrt{v(b)}}, \,\,\,\tilde{z} = \frac{z}{\sqrt{w(b)}},\,\,\, \mathcal{B}=\sqrt{3}B=\sqrt{3}\frac{b}{v(b)},\nonumber\\
\tilde{g}_{tt}&=-\frac{1}{\tilde{g}_{rr}}=-\tilde{U}(\tilde{r})=-U(r),\,\,\, \tilde{g}_{xx}=\tilde{g}_{yy}=e^{2\tilde{V}(\tilde{r})}=\frac{e^{2V(r)}}{v(b)},\,\,\, \tilde{g}_{zz}=e^{2\tilde{W}(\tilde{r})}=\frac{e^{2W(r)}}{w(b)}.
\label{eq:NumCoord}
\end{align}

It is important to remark that the aforementioned limitation on the range for the initial condition $b$, $b\in[0,\sqrt{3})$, does not imply in any limitation for the range of the physical dimensionless combination $\mathcal{B}/T^2=16\pi^2\sqrt{3}b/v(b)\in[0,\infty)$, because the function $v(b)$ monotonically decreases to zero as $b\to\sqrt{3}$.\footnote{As discussed in the introduction, the magnetic brane model has no dynamical breaking of the conformal symmetry, which is just explicitly broken by the magnetic field. This is the reason why the observables in this model are not functions of $\mathcal{B}$ and $T$ independently, but just of the dimensionless combination $\mathcal{B}/T^2$ (or any power of it).}

\subsection{Anisotropic jet quenching parameters}
\label{secDKjets}

With the numerical background functions determined as in Eqs. \eqref{eq:NumCoord}, one can plug them into the general formulas for the anistropic jet quenching parameters derived in section \ref{secANISO}. Below I explicitly write down the numerical integrations that need to be performed in terms of the numerical background functions of the magnetic brane model,
\begin{align}
\frac{\hat{q}_{\parallel(\perp)}}{\sqrt{\lambda_t}T^3} &= \frac{64\pi^2}{\int_{r_{\textrm{start}}}^{r_\textrm{max}} dr\,\frac{e^{-2V(r)}v(b)}{\sqrt{U(r)\left[e^{2W(r)}/w(b)-U(r)\right]}}},\\
\frac{\hat{q}_{\perp(\parallel)}}{\sqrt{\lambda_t}T^3} &= \frac{32\pi^2}{\int_{r_{\textrm{start}}}^{r_\textrm{max}} dr\,\frac{e^{-2W(r)}w(b)}{\sqrt{U(r)\left[e^{2V(r)}/v(b)-U(r)\right]}}},\\
\frac{\hat{q}_{\perp(\perp)}}{\sqrt{\lambda_t}T^3} &= \frac{32\pi^2}{\int_{r_{\textrm{start}}}^{r_\textrm{max}} dr\,\frac{e^{-2V(r)}v(b)}{\sqrt{U(r)\left[e^{2V(r)}/v(b)-U(r)\right]}}}.
\label{eq:DKjets}
\end{align}
An important remark on the numerical evaluation of the above integrals is the following. Due to numerical roundoff errors, for some values of the initial condition $b$ the terms between brackets inside the square roots may evaluate to small negative values at some large values of the radial coordinate (close to the boundary), artificially producing spurious complex results with a small imaginary part (compared to the real part). To avoid this issue, for each value of the initial condition $b$ the numerical routine I implemented searched for the first value of $r$ where a change of sign in the aforementioned terms happened within some stepsize precision, and in the cases where a change of sign was detected by the routine, then the upper limit of the numerical integration was cut off a bit before the value of $r$ corresponding to the onset of the region with undesirable roundoff errors.

\begin{figure}
\begin{center}
\begin{tabular}{c}
\includegraphics[width=0.5\textwidth]{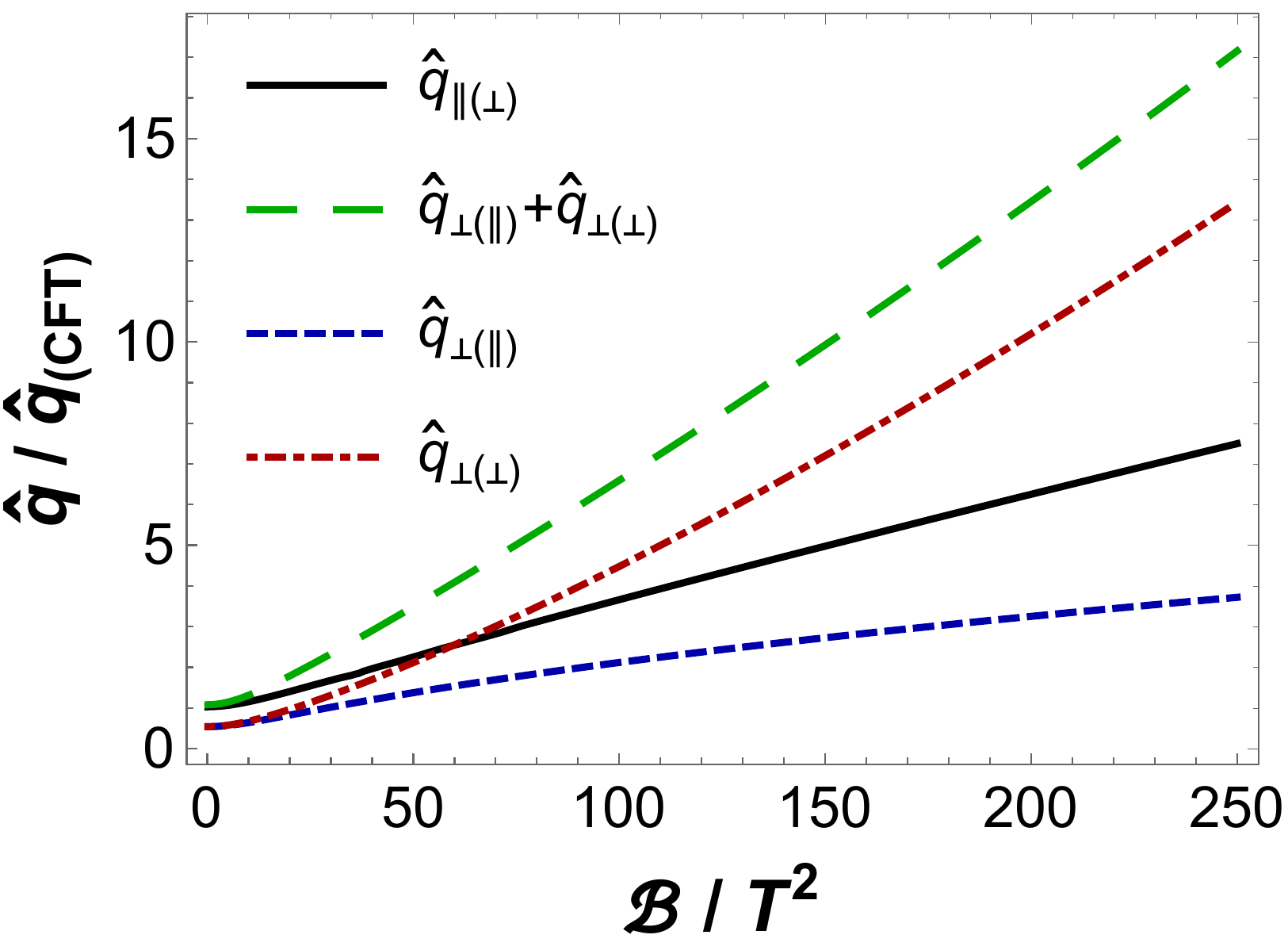} % \\
\end{tabular}
\end{center}
\caption{(Color online) Anisotropic jet quenching parameters for a light parton in the magnetic brane model normalized by the isotropic SYM result at zero magnetic field (conformal limit).}
\label{fig1}
\end{figure}

In the limit of vanishing anisotropy, $\mathcal{B}/T^2\to 0$, the anisotropic jet quenching parameters satisfy the constraint \eqref{eq:constraint}. It happens that for the magnetic brane model this is also the conformal limit ($T\gg\sqrt{\mathcal{B}}$), corresponding to the result for the SYM plasma \cite{Liu:2006ug},
\begin{align}
\frac{\hat{q}_{(\textrm{CFT})}}{\sqrt{\lambda_t}T^3}=\frac{\pi^{3/2}\Gamma(3/4)}{\Gamma(5/4)}.
\label{eq:CFT}
\end{align}
The full numerical results for the anisotropic jet quenching parameters of the magnetic brane model normalized by the above conformal limit are shown in Fig. \ref{fig1}.

One concludes that for the magnetic brane model all the jet quenching parameters monotonically increase with increasing $\mathcal{B}/T^2$ and, furthermore,
\begin{align}
\hat{q}_{\perp(\parallel)}+\hat{q}_{\perp(\perp)}&\ge\hat{q}_{\parallel(\perp)}\ge \hat{q}_{(\textrm{iso})}=\hat{q}_{(\textrm{CFT})},\label{eq:hierarchyDK1}\\
\hat{q}_{\perp(\perp)}&\ge\hat{q}_{\perp(\parallel)},\label{eq:hierarchyDK2}
\end{align}
with the equalities being saturated in the limit of zero magnetic field. In Ref. \cite{Li:2016bbh}, working with an analytical approximation for the magnetic brane background strictly valid in the limit of strong magnetic fields, $\mathcal{B}\gg T^2$, it was concluded that $\hat{q}_{\perp(\perp)}>\hat{q}_{\perp(\parallel)}$. From Fig. \ref{fig1} I have shown that this result is indeed valid for any finite value of $\mathcal{B}/T^2$.

\section{The magnetic EMD model}
\label{secEMD}

\subsection{The holographic model}
\label{secEMDmodel}

In this section I briefly review the basics of the phenomenological magnetic EMD model \cite{Finazzo:2016mhm}, regarding what is just necessary for the computation of the corresponding numerical backgrounds and the anisotropic jet quenching parameters.

The bulk action for the EMD model reads,
\begin{align}
S&=\frac{1}{16\pi G_5}\int_{\mathcal{M}_5}d^5x\sqrt{-g}\left[R-\frac{1}{2}(\partial_\mu\phi)^2-V(\phi) -\frac{f(\phi)}{4}F_{\mu\nu}^2\right],
\label{eqSS}
\end{align}
which is supplemented by boundary terms related to the holographic renormalization of the model, as before. The free parameters of this bottom-up construction are dynamically fixed by $(2+1)$-flavors lattice QCD inputs with physical quark masses. These inputs are the QCD equation of state \cite{Borsanyi:2013bia} and the magnetic susceptibility \cite{Bonati:2013vba}, both computed at zero magnetic field. Dimensionful observables in the dual gauge theory at the boundary are naturally measured in inverse powers of the asymptotic AdS$_5$ radius, which was set to unity as before. In order to express these observables in physical units, it is introduced a fixed scaling factor $\Lambda$ [MeV], such that any physical observable $X$ at the boundary quantum field theory with mass dimension $q$ is expressed in physical units as $X=\hat{X}\Lambda^q$ [MeV$^q$], where $\hat{X}$ denotes the observable calculated in the bulk gravity theory in units of inverse AdS$_5$ radius.\footnote{Notice this procedure does not introduce any extra free parameter in the bulk EMD action, since it just amounts to exchange the freedom of fixing the value of the asymptotic AdS$_5$ radius with the freedom to choose the value of the scaling parameter $\Lambda$.} In Ref. \cite{Finazzo:2016mhm} there were fixed in this way the following set of model parameters,
\begin{align}
V(\phi)&=-12\cosh(0.63\phi)+0.65\phi^2-0.05\phi^4+0.003\phi^6,\nonumber\\
\kappa^2&=8\pi G_5=8\pi(0.46), \quad \Lambda=1058.83\,\textrm{MeV},\nonumber\\
f(\phi)&=0.95\,\textrm{sech}(0.22\phi^2-0.15\phi-0.32),
\label{eq:fits}
\end{align}
where from the dilaton potential above one obtains that the scaling dimension of the dual relevant operator in the boundary gauge theory is $\Delta\equiv 4-\nu\approx 2.73$.

Some general remarks regarding the nature of the scalar field $\phi$ in this bottom-up EMD model are in order at this point. The effective dilaton potential $V(\phi)$ specified above implies a 5D massive scalar field, and one can ask whether this scalar field is indeed the dilaton (which is usually massless in 10D), or some other scalar. In this regard, there are some processes, like SUSY breaking, which gives mass to the dilaton. One other possibility is that this is a massive 5D KK mode of the massless 10D dilaton obtained after some specific 5D compactification. Nevertheless, since this is a bottom-up construction, the origin of this mass in the EMD action is currently unknown. By the same token, since the dual QFT at the boundary is unknown in bottom-up constructions, one cannot say for sure to which specific scalar operator this scalar field couples at the boundary, although one knows that its scaling dimension is $\Delta\approx 2.73$, as implied by $V(\phi)$ (which, in turn, was phenomenologically fixed by matching the lattice QCD equation of state at zero magnetic field). Of course, it is also possible that $\phi$ is other scalar field unrelated to the dilaton. In previous works involving bottom-up Einstein-scalar models, as for instance in Refs. \cite{Critelli:2016cvq,DeWolfe:2009vs,Ewerz:2016zsx}, the effects of considering $\phi$ to be or not the dilaton were considered for different observables.\footnote{If $\phi$ is unrelated to the dilaton, then the string and Einstein frames coincide.} Generally, the results are drastically different, even at the qualitative level, depending on the nature of this scalar field. Of particular relevance to the present work is the result of Ref. \cite{Critelli:2016cvq}, where by considering $\phi$ as being the dilaton, a good quantitative agreement was found between the EMD predictions for the Polyakov loop and specially the heavy quark entropy in the deconfined plasma, and the corresponding lattice QCD results. On the other hand, by considering instead $\phi$ to be some other scalar field, the holographic results obtained for the aforementioned observables have nothing to do with the corresponding lattice results, even qualitatively. Even though these facts do not constitute a formal proof that $\phi$ in this model is the dilaton, they strongly favor this possibility. In particular, for practical applications, in face of the aforementioned results, it seems that the only phenomenologically viable approach is to consider $\phi$ as being the dilaton, what is done in the present work.\footnote{It is also important to comment that in the general results displayed in section \ref{secANISO} the usual coupling $\phi\mathcal{R}$ between the dilaton and the 2D Ricci scalar induced on the string worldsheet was neglected, because the working hypothesis assumed here considers the classical gauge/gravity limit of the holographic duality, in which the 't Hooft coupling $\lambda_t$ is large. Since the term $\phi\mathcal{R}$ is of order zero in $\lambda_t$, it is negligible when compared to the Nambu-Goto action which is of order $1/2$. On the other hand, if one considers that the 't Hooft coupling is not large, then one should not only consider the $\phi\mathcal{R}$ term in the worldsheet action, but also higher order derivative corrections of the metric field in the bulk action. I do not consider such finite 't Hooft coupling corrections in the present work.}

\newpage

A final remark concerning the dilaton potential regards the fact that, as mentioned below Eq. \eqref{eqSS}, this potential was fixed here by using inputs from lattice QCD simulations with $(2+1)$-flavors and physical quark masses. The standard way conveyed by the holographic dictionary to take into account the flavor dynamics of the boundary gauge theory on the gravity side of the gauge/gravity duality is by means of the introduction of flavor branes within the bulk. Here I followed another approach, pionereed by Gubser and collaborators in Refs. \cite{Gubser:2008ny,Gubser:2008yx,Gubser:2008sz,DeWolfe:2010he,DeWolfe:2011ts}, where the dilaton potential is assumed to effectively encode both the dynamics of the flavor sector as well as the dynamics of the gluonic sector. This is the reason why the dilaton potential was constructed here by directly using lattice QCD data with flavors. The phenomenological reliability of such approach can be checked in practice by contrasting the model predictions with the corresponding first principles lattice QCD data (when available). As discussed in the introduction, the present EMD model has been verified in previous works \cite{Finazzo:2016mhm,Critelli:2016cvq} to quantitatively predict a variety of lattice QCD data (which were not used to fix the dilaton potential), thus providing strong evidence that such approach is completely feasible for phenomenological applications in QCD.

The ansatz for the anisotropic EMD magnetic backgrounds in the standard coordinates is given by,
\begin{align}
d\tilde{s}^2&=e^{2\tilde{a}(\tilde{r})}\left[-\tilde{h}(\tilde{r})d\tilde{t}^2+d\tilde{z}^2\right]+ e^{2\tilde{c}(\tilde{r})}(d\tilde{x}^2+d\tilde{y}^2)+\frac{e^{2\tilde{b}(\tilde{r})} d\tilde{r}^2}{\tilde{h}(\tilde{r})},\nonumber\\
\tilde{\phi}&=\tilde{\phi}(\tilde{r}),\,\,\, \tilde{A}=\tilde{A}_\mu d\tilde{x}^\mu=\hat{B}\tilde{x}d\tilde{y}\Rightarrow \tilde{F}=d\tilde{A}=\hat{B}d\tilde{x}\wedge d\tilde{y}.
\label{2.14}
\end{align}
Maxwell's equations are trivially satisfied by the ansatz \eqref{2.14} and the equation of motion for the dilaton field reads,
\begin{align}
\tilde{\phi}'' + \left(2\tilde{a}'+2\tilde{c}'-\tilde{b}'+\frac{\tilde{h}'}{\tilde{h}} \right) \tilde{\phi}' - \frac{e^{2\tilde{b}}}{\tilde{h}} \left(\frac{\partial V}{\partial\tilde{\phi}} + \frac{\hat{B}^2 e^{-4\tilde{c}}}{2} \frac{\partial f}{\partial\tilde{\phi}} \right) = 0,
\label{eq:dil}
\end{align}
while Einstein's equations can be worked out to give,
\begin{align}
\tilde{a}''+\left( \frac{14}{3} \tilde{c}' - \tilde{b}' + \frac{4}{3} \frac{\tilde{h}'}{h} \right) \tilde{a}' + \frac{8}{3} \tilde{a}'^2 + \frac{2}{3} \tilde{c}'^2 + \frac{2}{3} \frac{\tilde{h}'}{\tilde{h}} \tilde{c}' + \frac{2}{3} \frac{e^{2\tilde{b}}}{\tilde{h}} V - \frac{1}{6} \tilde{\phi}'^2 & =0, \label{eq:a}\\
\tilde{c}'' - \left(\frac{10}{3} \tilde{a}'+\tilde{b}'+\frac{1}{3} \frac{\tilde{h}'}{\tilde{h}} \right) + \frac{2}{3} \tilde{c}'^2 - \frac{4}{3} \tilde{a}'^2 - \frac{2}{3} \frac{\tilde{h}'}{\tilde{h}} \tilde{a}' - \frac{1}{3}\frac{e^{2\tilde{b}}}{\tilde{h}} V + \frac{1}{3} \tilde{\phi}'^2 & = 0, \label{eq:c}\\
\tilde{h}''+(2\tilde{a}'+2\tilde{c}'-\tilde{b}')\tilde{h}' & = 0. \label{eq:h}
\end{align}
One can derive from the above equations the following constraint on horizon data,
\begin{equation}
\tilde{a}'^2 + \tilde{c}'^2 - \frac{1}{4} \tilde{\phi}'^2 + \left( \frac{\tilde{a}'}{2} + \tilde{c}' \right)\frac{\tilde{h}'}{\tilde{h}} + 4\tilde{a}'\tilde{c}' + \frac{e^{2\tilde{b}}}{2\tilde{h}} \left( V + \frac{\hat{B}^2 e^{-4\tilde{c}}}{2} f \right) = 0.
\label{eqcons}
\end{equation}
The background function $\tilde{b}(\tilde{r})$ has no equation of motion to satisfy and can be set to zero, $\tilde{b}(\tilde{r})=0$.

In order to ascribe numerical values to all the horizon data required to initialize the numerical routine to integrate the coupled ordinary differential equations \eqref{eq:dil} -- \eqref{eq:h}, one specifies the numerical coordinates as follows.
First, one writes down the ansatz for the bulk fields in these coordinates,
\begin{align}
ds^2&=e^{2a(r)}\left[-h(r)dt^2+dz^2\right]+e^{2c(r)}(dx^2+dy^2)+\frac{dr^2}{h(r)},\nonumber\\
\phi&=\phi(r),\,\,\,A=A_\mu dx^\mu=\mathcal{B}xdy\Rightarrow F=dA=\mathcal{B}dx\wedge dy.
\end{align}
Let now $Y(r)\in\left\{a(r), c(r), h(r), \phi(r)\right\}$. By Taylor expanding these background functions near the horizon to second order,
\begin{equation}
Y(r) = Y_0 + Y_1(r-r_H) + Y_2 (r-r_H)^2 + \ldots,
\end{equation}
one needs to specify 12 infrared Taylor expansion coefficients to start the numerical integration of the equations of motion. One of them is the horizon value of the dilaton field, $\phi_0$, which corresponds to one of the initial conditions of the system. The other initial condition is the value of the magnetic field in the numerical coordinates, $\mathcal{B}$. As detailed discussed in Ref. \cite{Rougemont:2015oea}, by rescaling the bulk coordinates one may fix, $r_H = 0$, $a_0 = c_0 = 0$, and $h_1 = 1$, with $h_0 = 0$ following from the definition of the background blackening function $h(r)$. The 7 remaining infrared coefficients can be dynamically fixed as functions of the initial conditions $(\phi_0,\mathcal{B})$ by substituting the infrared expansions back into Eqs. \eqref{eq:dil} -- \eqref{eqcons} and then solving the resulting algebraic system.

With the horizon data specified as above, one can now numerically integrate the equations of motion for different values of the initial conditions $(\phi_0,\mathcal{B})$. As before, in order to avoid the singular point of the differential equations at the horizon, one starts the numerical integration procedure slightly beyond it, at for instance, $r_{\textrm{start}} = 10^{-8}$. The boundary is formally at infinity, but for numerical purposes one clearly needs to stop it at some finite value of $r$, where the ultraviolet fixed point corresponding to the AdS$_5$ geometry is reached. For the set of initial conditions considered in the present work to plot the physical observables within the region $T\,=\,130$ -- $400$ MeV and $eB\,=0\,$ -- $0.6$ GeV$^2$, it is enough to work with initial conditions within the ranges, $\phi_0\in[0.3,4.0]$ and $\mathcal{B}/\mathcal{B}_{\textrm{max}}(\phi_0)\in[0,0.9]$, where the meaning of $\mathcal{B}_{\textrm{max}}(\phi_0)$ is going to be discussed in a moment. For this range of initial conditions, stopping the numerical integration of the equations of motion at $r_{\textrm{max}} = 2$ warrants that the generated numerical backgrounds reach AdS$_5$ at (or generally before) this ultraviolet cutoff. For each chosen value of the pair of initial conditions $(\phi_0,\mathcal{B})$ there is generated a numerical solution corresponding to some specific physical state at the dual boundary gauge theory.

As in the previous case of the magnetic brane model, also in the magnetic EMD model there is an upper bound on the values of the initial condition $\mathcal{B}$ for which asymptotically AdS$_5$ geometries can be generated. In this case, this upper bound is a function of the first initial condition, $\phi_0$, and it is denoted here by $\mathcal{B}_{\textrm{max}}(\phi_0)$. This bound can be numerically determined as discussed in Ref. \cite{Finazzo:2016mhm}.\footnote{For the derivation of an analogous analytical bound in the context of the isotropic EMD model at finite temperature and baryon chemical potential, see Ref. \cite{DeWolfe:2010he}.}

For the observables plotted in the present work I generated a regularly spaced $150\times 150$ rectangular grid of initial conditions within the aforementioned ranges for $\phi_0$ and $\mathcal{B}$. Within the physical region analyzed in the present work, corresponding to $T\,=\,130$ -- $400$ MeV and $eB\,=0\,$ -- $0.6$ GeV$^2$, no actual phase transition is observed (just a smooth crossover, as in lattice QCD simulations \cite{Bali:2014kia}). With just $22,500$ generated points irregularly spaced in the $(T,eB)$ plane,\footnote{The mapping of the initial conditions to the physical space of temperature and magnetic field distorts a regularly spaced rectangle in the $(\phi_0,\mathcal{B})$ plane into an irregularly shaped region with highly asymmetric distribution of points in the $(T,eB)$ plane. One possible way of augmenting the uniformity in the distribution of points within the $(T,eB)$ plane is by randomly generating the initial conditions, instead of using fixed stepsizes in the $\phi_0$ and $\mathcal{B}$ directions.} the numerical interpolations done as functions of $T$ and $eB$ present some small oscillations. By largely augmenting the number of generated points within the fixed ranges of the initial conditions these oscillations are eliminated and the numerical interpolations are smoother. However, this procedure also largely increases the computation time.

\newpage

A much faster alternative to generate smoother interpolations (although not so efficient in smoothing out the aforementioned numerical oscillations) is to reinterpolate the originally interpolated background functions in terms of a finer grid with the same boundaries. In the plots presented in Figs. \ref{fig2} and \ref{fig3} I employed this reinterpolation procedure on top of a finer evenly spaced $600\times 600$ grid. The numerical oscillations are not clearly seen in the 2D plots, but they can be noticed by the spots in the 3D plots.

As before, in order to use standard holographic formulas for the physical observables at the boundary gauge theory one needs to relate the numerical coordinates, where the numerical solutions are obtained, with the standard (tilded) coordinates. This has been done in details in Refs. \cite{Finazzo:2016mhm,Rougemont:2015oea}, and I summarize below the corresponding results,
\begin{align}
\tilde{r}&=\frac{r}{\sqrt{h_0^{\textrm{far}}}}+a_0^{\textrm{far}}-\ln\left(\phi_A^{1/\nu}\right),\,\,\,
\tilde{t}=\phi_A^{1/\nu}\sqrt{h_0^{\textrm{far}}}t,\,\,\,
\tilde{x}=\phi_A^{1/\nu}e^{c_0^{\textrm{far}}-a_0^{\textrm{far}}}x,\nonumber\\
\tilde{y}&=\phi_A^{1/\nu}e^{c_0^{\textrm{far}}-a_0^{\textrm{far}}}y,\,\,\,
\tilde{z}=\phi_A^{1/\nu}z,\,\,\,
\tilde{a}(\tilde{r})=a(r)-\ln\left(\phi_A^{1/\nu}\right),\nonumber\\
\tilde{c}(\tilde{r})&=c(r)-(c_0^{\textrm{far}}-a_0^{\textrm{far}})-\ln\left(\phi_A^{1/\nu}\right),\,\,\,
\tilde{h}(\tilde{r})=\frac{h(r)}{h_0^{\textrm{far}}},\,\,\,
\tilde{\phi}(\tilde{r})=\phi(r),\nonumber\\
e\hat{B}&=\frac{e^{2(a_0^{\textrm{far}}-c_0^{\textrm{far}})}}{\phi_A^{2/\nu}}\mathcal{B},\nonumber\\
\hat{T}&= \frac{1}{4\pi \phi_A^{1/\nu} \sqrt{h^{\mathrm{far}}_0}},
\label{eq:RelationsEMD-Stan-Num}
\end{align}
where the set of ultraviolet near-boundary coefficients extracted from the background functions in the numerical coordinates can be fixed as follows \cite{Finazzo:2016mhm}: $h_0^{\textrm{far}}=h(r_{\textrm{max}})$; $a^{\mathrm{far}}_0$ and $c^{\mathrm{far}}_0$ can be obtained by matching the numerical results with the ultraviolet fitting profiles $a(r)=a^{\mathrm{far}}_0+r/\sqrt{h_0^{\textrm{far}}}$ and $c(r)=c^{\mathrm{far}}_0+r/\sqrt{h_0^{\textrm{far}}}$ within the interval $r\in [1,r_{\textrm{max}}]$; $\phi_A$ can be extracted by first defining the adaptive variables, $r_{\textrm{IR}}(\phi_0,\mathcal{B})\equiv \phi^{-1}(10^{-3})$ and $r_{\textrm{UV}}(\phi_0,\mathcal{B})\equiv \phi^{-1}(10^{-5})$, and then matching the numerical results with the ultraviolet fitting profile $\phi(r)=\phi_A e^{-\nu a(r)}$ within the adaptive interval $r\in [r_{\textrm{IR}},r_{\textrm{UV}}]$.

This anisotropic magnetic EMD model was shown in Refs. \cite{Finazzo:2016mhm,Critelli:2016cvq} to be able to quantitatively \emph{predict} the finite temperature and magnetic field behavior of the $(2+1)$-flavors QCD equation of state with physical quark masses, the renormalized Polyakov loop, and the heavy quark entropy in the deconfined QGP phase up to the highest values of magnetic field currently reached in state-of-the-art lattice QCD simulations \cite{Bali:2014kia,Bruckmann:2013oba,Endrodi:2015oba,Bazavov:2016uvm}.

\subsection{Anisotropic jet quenching parameters}
\label{secEMDjets}

The background metric components in the Einstein frame are, according to Eqs. \eqref{2.14} and \eqref{eq:RelationsEMD-Stan-Num},
\begin{align}
\tilde{g}_{tt}&= -\tilde{h}(\tilde{r})e^{2\tilde{a}(\tilde{r})}= -\frac{h(r)}{h_0^{\textrm{far}}} \frac{e^{2a(r)}}{\phi_A^{2/\nu}},\nonumber\\
\tilde{g}_{rr}&= \frac{1}{\tilde{h}(\tilde{r})} = \frac{h_0^{\textrm{far}}}{h(r)},\nonumber\\
\tilde{g}_{xx}&=\tilde{g}_{yy}= e^{2\tilde{c}(\tilde{r})}= \frac{e^{2(c(r)-c_0^{\textrm{far}}+a_0^{\textrm{far}})}}{\phi_A^{2/\nu}},\nonumber\\
\tilde{g}_{zz}&= e^{2\tilde{a}(\tilde{r})}= \frac{e^{2a(r)}}{\phi_A^{2/\nu}}.
\label{eq:EMDtransf}
\end{align}
With this, I write down below the numerical integrations that need to be performed on top of the numerical magnetic EMD backgrounds in order to evaluate the anisotropic jet quenching parameters,

\begin{align}
\frac{\hat{q}_{\parallel(\perp)}}{\sqrt{\lambda_t}T^3} &= \frac{64\pi^2 h_0^{\textrm{far}}}{\int_{r_{\textrm{start}}}^{r_\textrm{max}} dr\,\frac{e^{-\sqrt{2/3}\,\phi(r)-2c(r)-a(r)+2\left(c_0^{\textrm{far}}-a_0^{\textrm{far}}\right)}}{ \sqrt{h(r)\left[h_0^{\textrm{far}}-h(r)\right]}}},\\
\frac{\hat{q}_{\perp(\parallel)}}{\sqrt{\lambda_t}T^3} &= \frac{32\pi^2 h_0^{\textrm{far}}}{\int_{r_{\textrm{start}}}^{r_\textrm{max}} dr\,\frac{e^{-\sqrt{2/3}\,\phi(r)-2a(r)}}{ \sqrt{h(r)\left[h_0^{\textrm{far}}e^{2c(r)-2\left( c_0^{\textrm{far}} - a_0^{\textrm{far}} \right)}-h(r)e^{2a(r)}\right]}}},\\
\frac{\hat{q}_{\perp(\perp)}}{\sqrt{\lambda_t}T^3} &= \frac{32\pi^2 h_0^{\textrm{far}}}{\int_{r_{\textrm{start}}}^{r_\textrm{max}} dr\,\frac{e^{-\sqrt{2/3}\,\phi(r)-2c(r)+2(c_0^{\textrm{far}}-a_0^{\textrm{far}})}}{ \sqrt{h(r)\left[h_0^{\textrm{far}}e^{2c(r)-2\left( c_0^{\textrm{far}} - a_0^{\textrm{far}} \right)}-h(r)e^{2a(r)}\right]}}}.
\label{eq:EMDjets}
\end{align}
The same observation done below Eq. \eqref{eq:DKjets} regarding roundoff errors in the numerical integrations for the anisotropic jet quenching parameters also applies here.

In the isotropic limit of zero magnetic field, $c(r)=a(r)$ and $c_0^{\textrm{far}}=a_0^{\textrm{far}}$, therefore it follows from the above formulas that the constraint \eqref{eq:constraint} is satisfied and the formula for the isotropic jet quenching parameter, $\hat{q}_{(\textrm{iso})}$, originally derived in Ref. \cite{Rougemont:2015wca} for the isotropic EMD model is recovered.

\begin{figure}
\begin{center}
\begin{tabular}{c}
\includegraphics[width=0.45\textwidth]{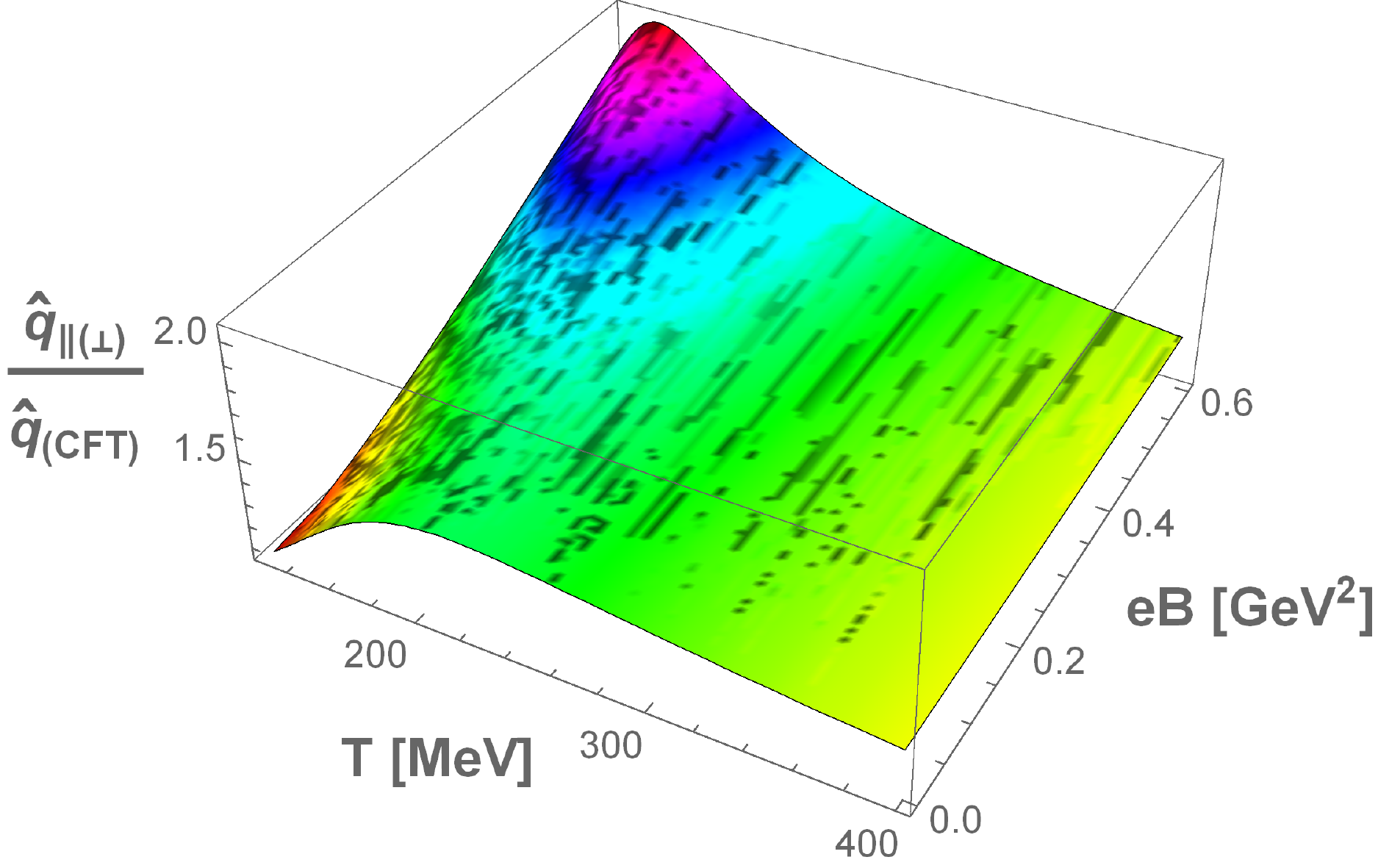} % \\
\end{tabular}
\begin{tabular}{c}
\includegraphics[width=0.45\textwidth]{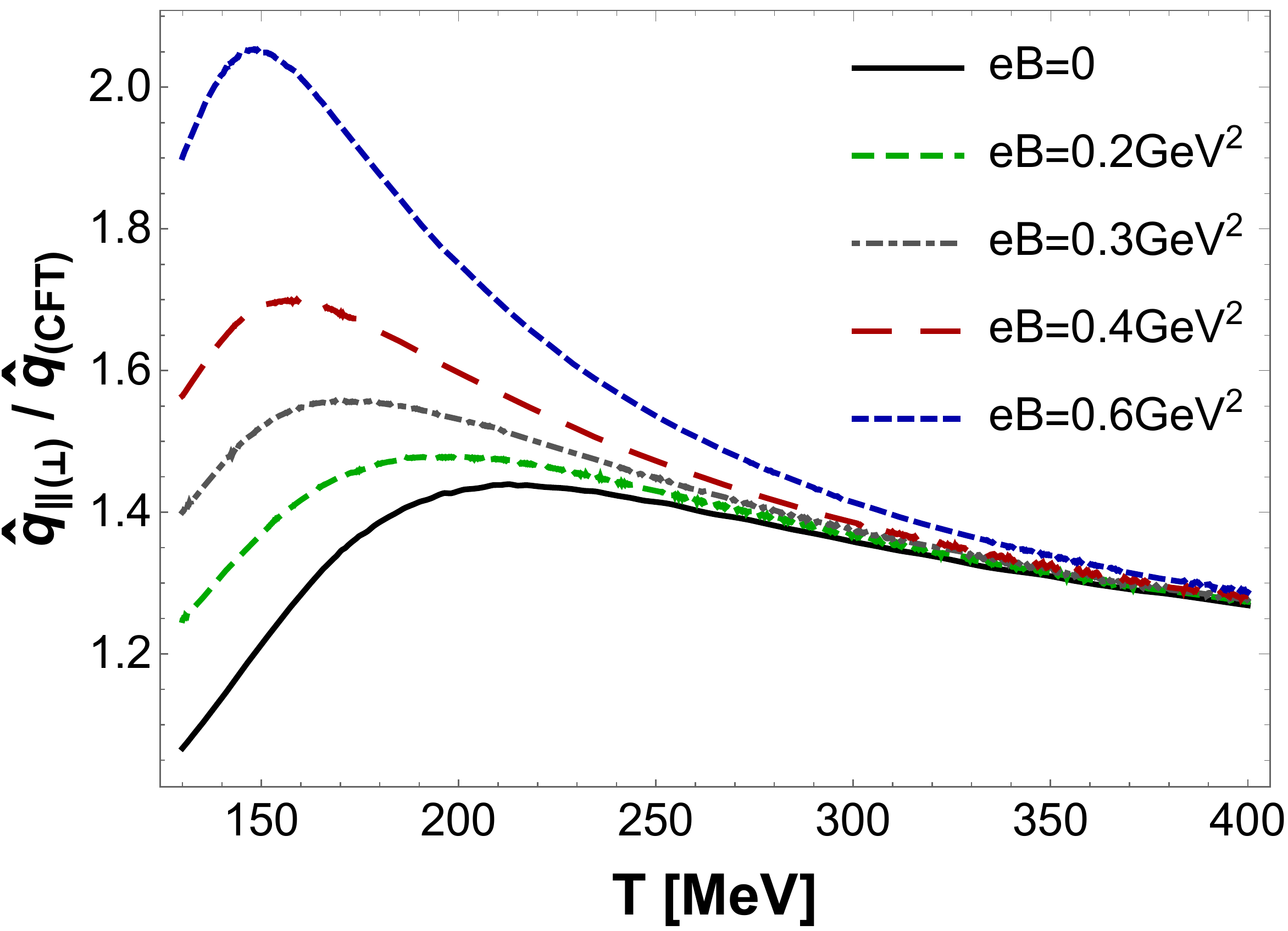} % \\
\end{tabular}
\end{center}
\caption{(Color online) Anisotropic jet quenching parameter for a light parton moving parallel to the magnetic field in the magnetic EMD model normalized by the isotropic SYM result (conformal limit).}
\label{fig2}
\end{figure}

\begin{figure}
\begin{center}
\begin{tabular}{c}
\includegraphics[width=0.45\textwidth]{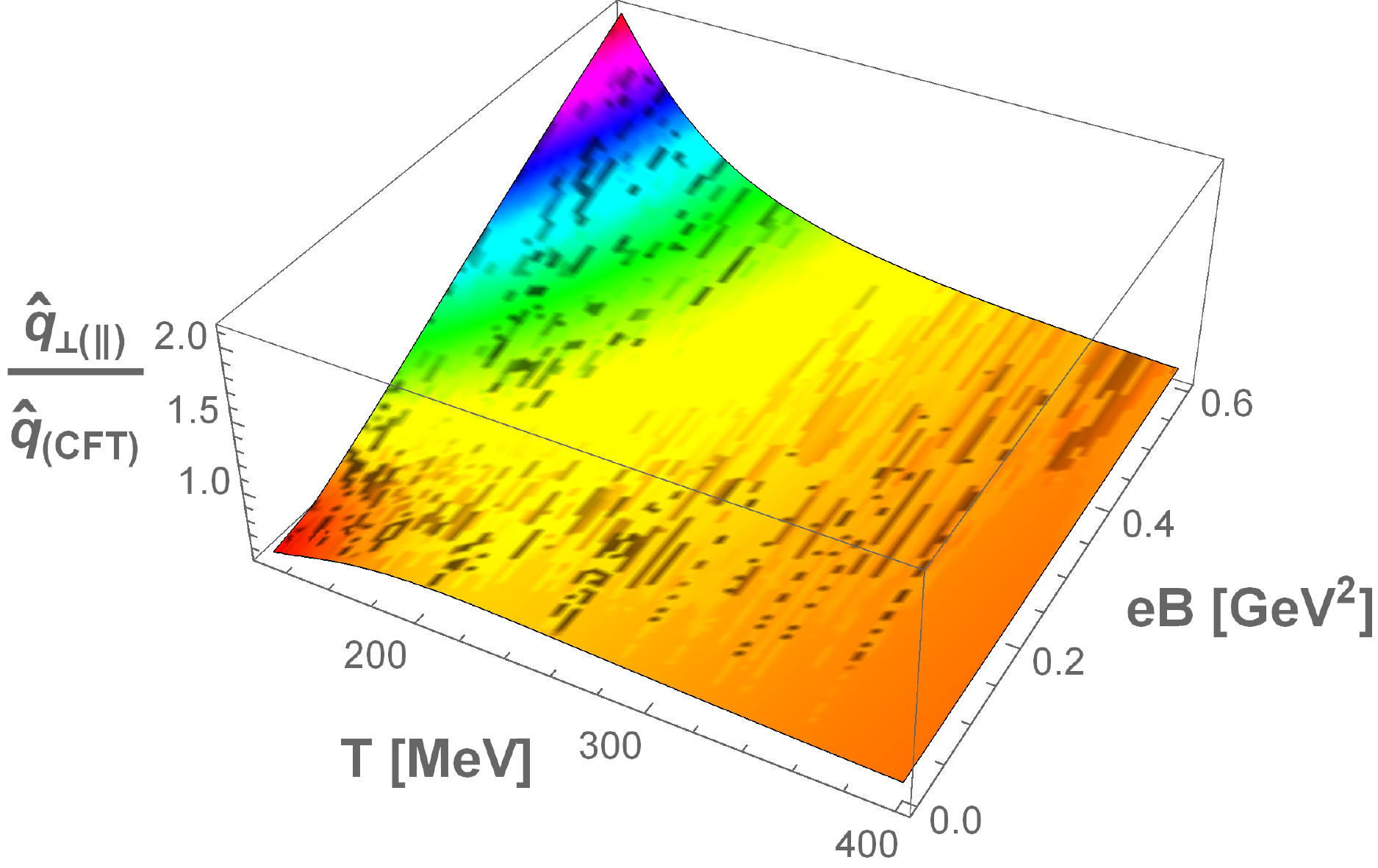} % \\
\end{tabular}
\begin{tabular}{c}
\includegraphics[width=0.45\textwidth]{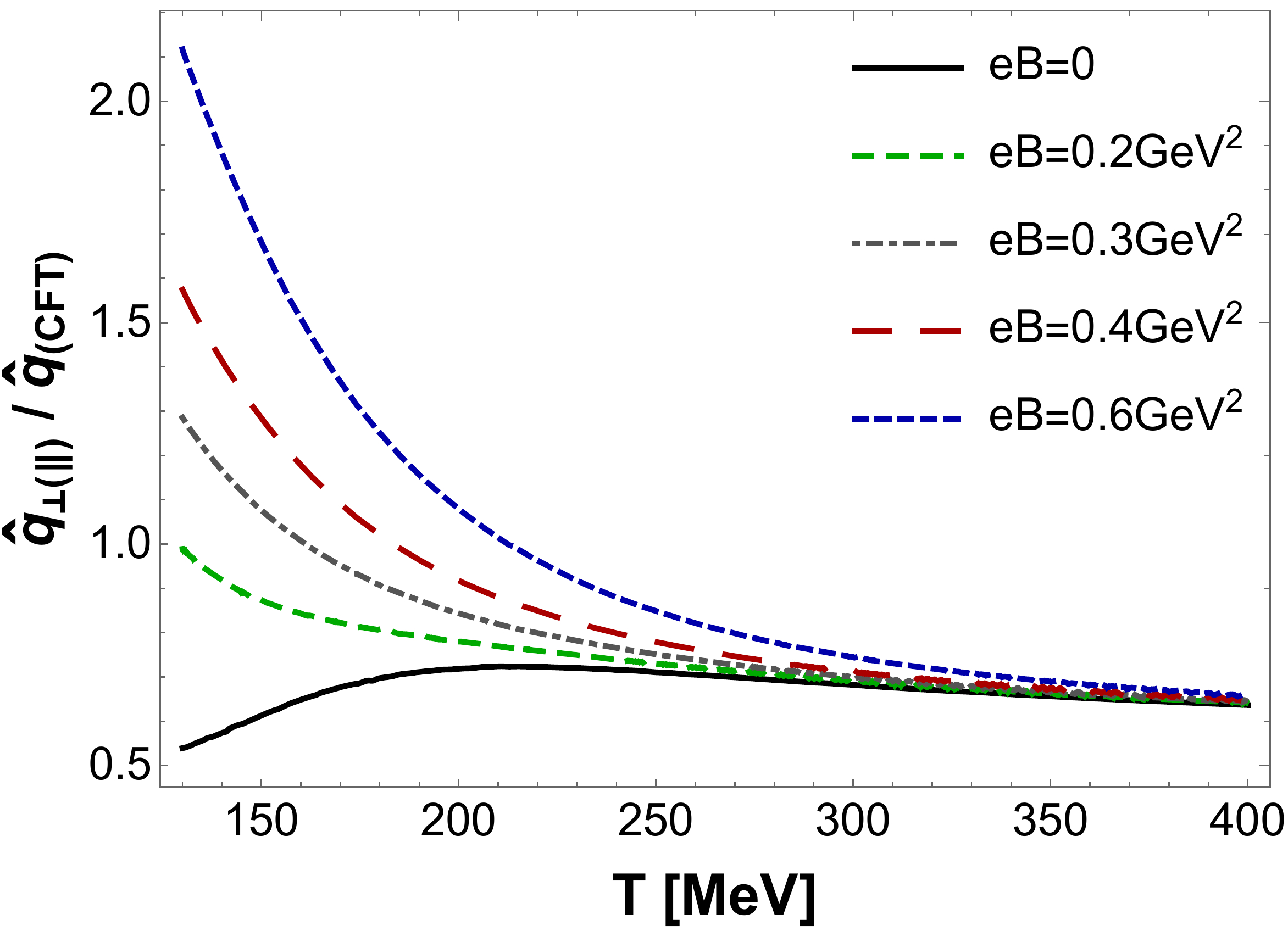} % \\
\end{tabular}
\end{center}
\begin{center}
\begin{tabular}{c}
\includegraphics[width=0.45\textwidth]{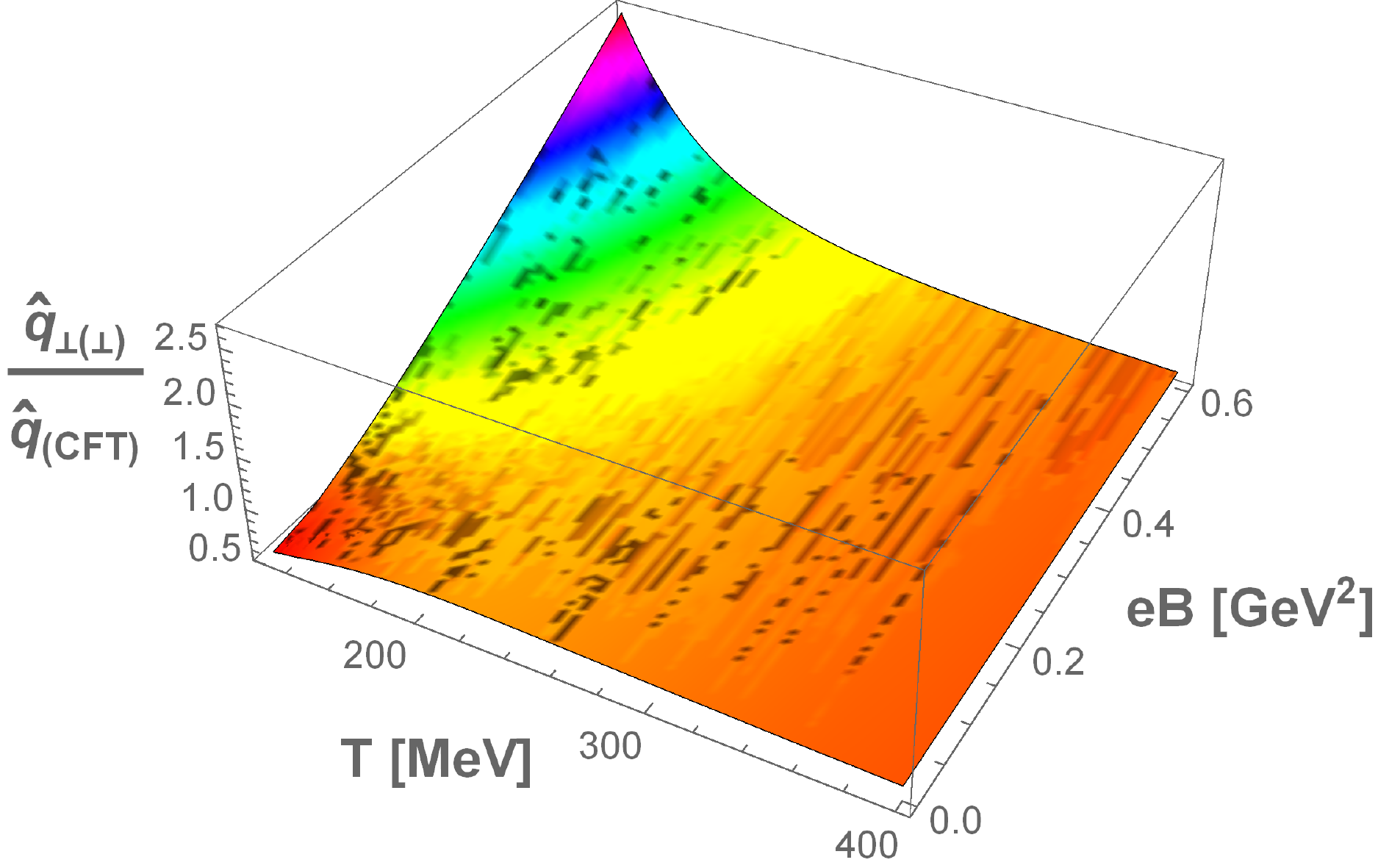} % \\
\end{tabular}
\begin{tabular}{c}
\includegraphics[width=0.45\textwidth]{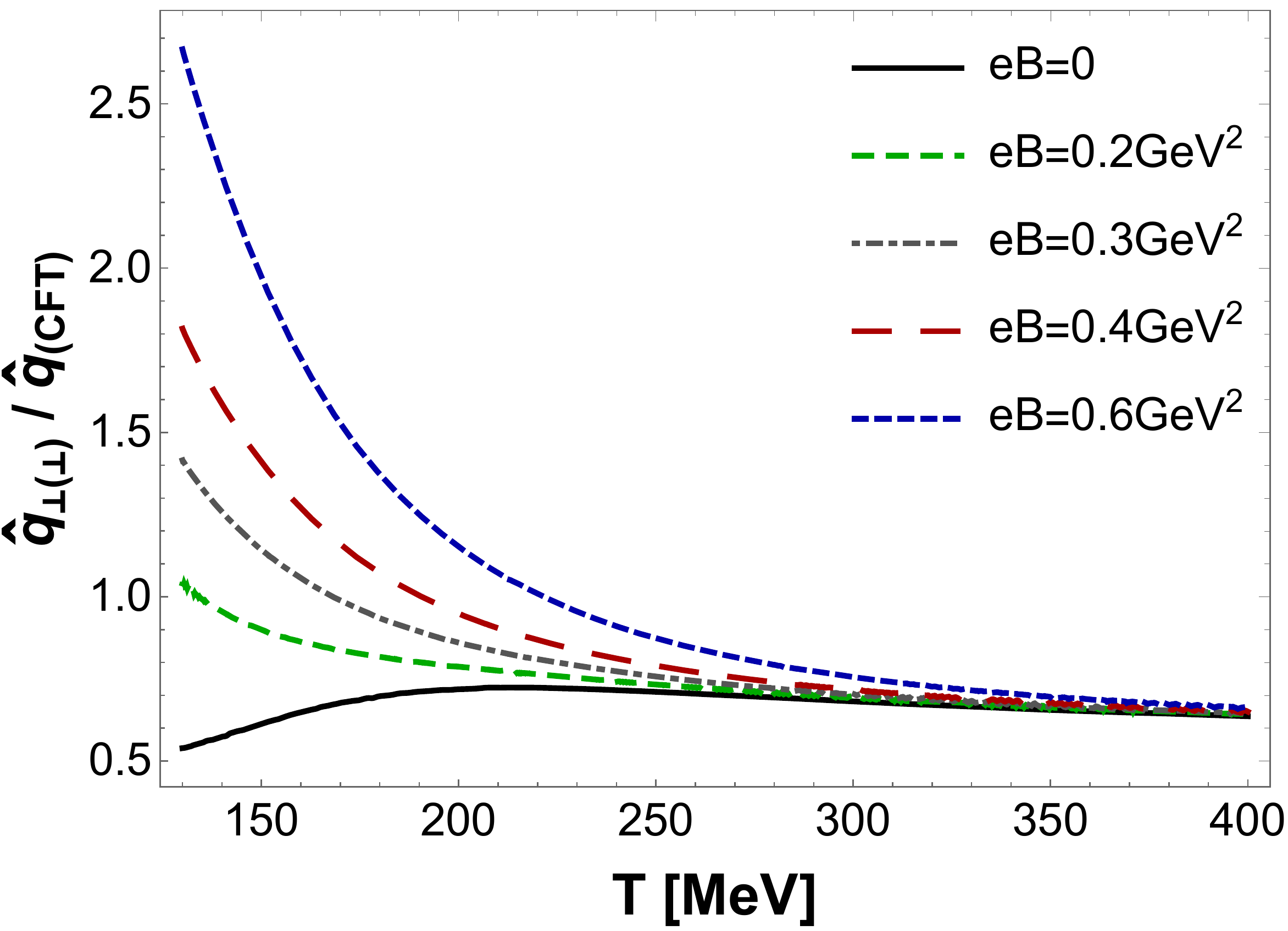} % \\
\end{tabular}
\end{center}
\begin{center}
\begin{tabular}{c}
\includegraphics[width=0.45\textwidth]{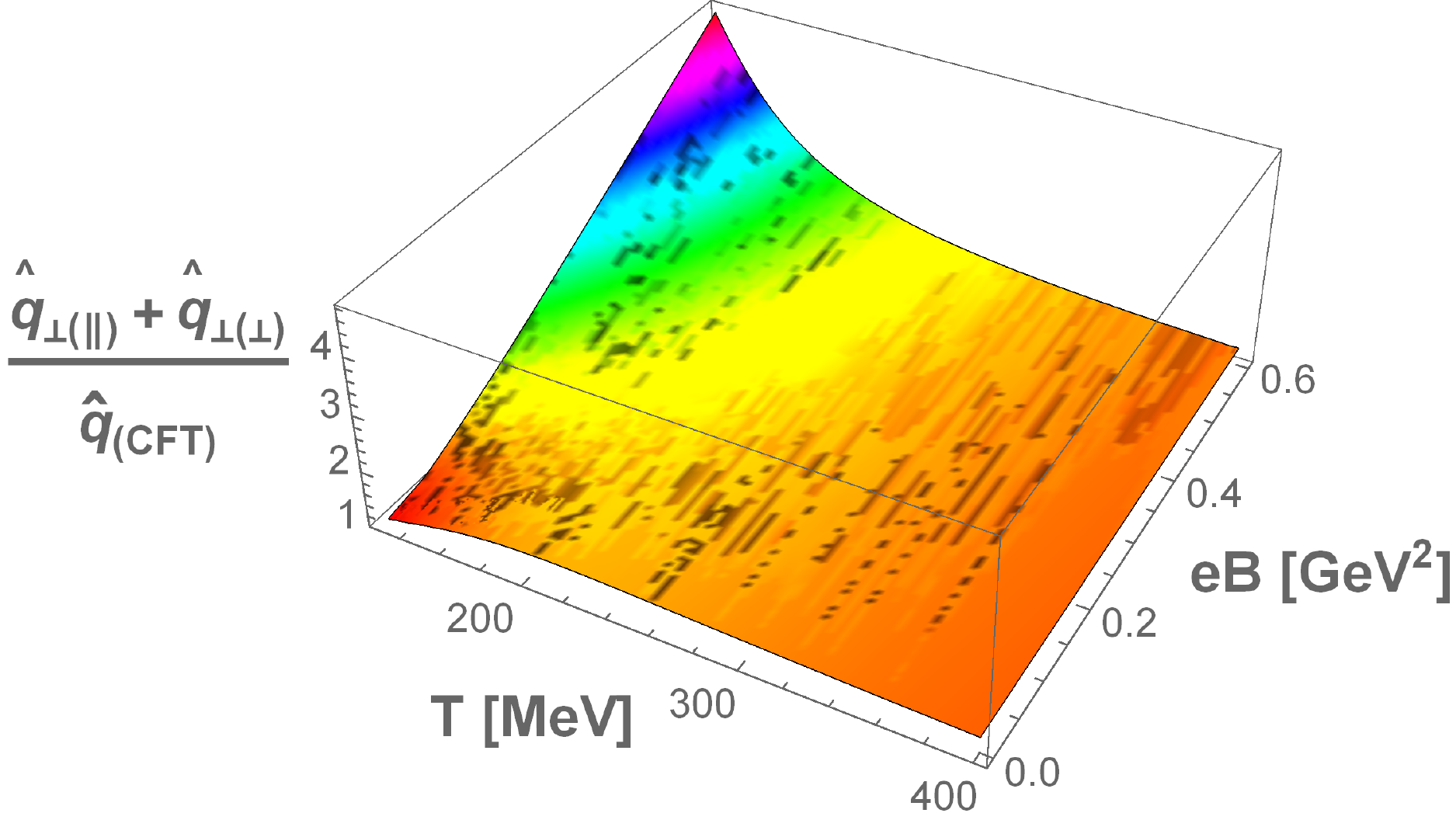} % \\
\end{tabular}
\begin{tabular}{c}
\includegraphics[width=0.45\textwidth]{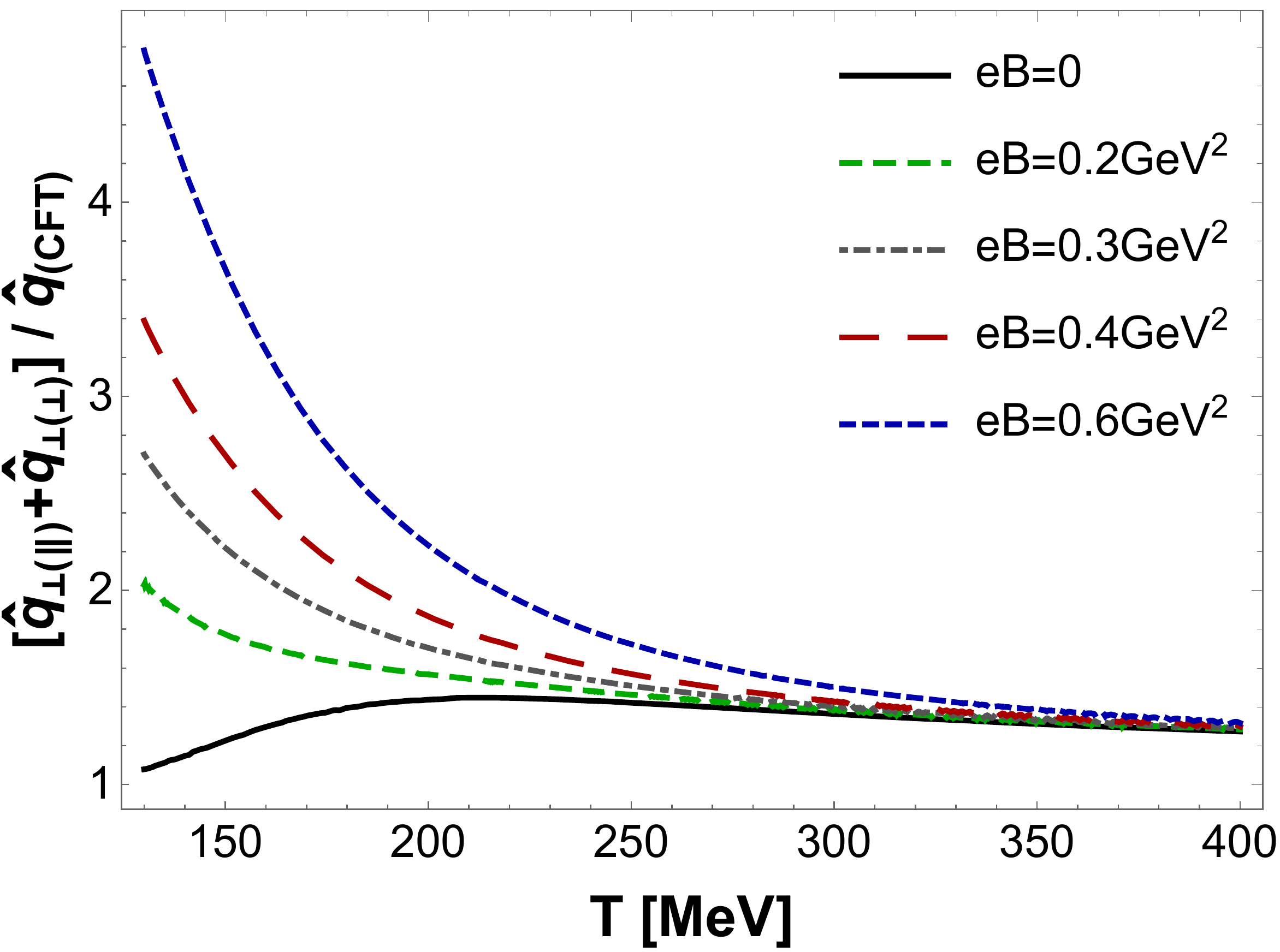} % \\
\end{tabular}
\end{center}
\caption{(Color online) Anisotropic jet quenching parameters for a light parton moving perpendicular to the magnetic field in the magnetic EMD model normalized by the isotropic SYM result (conformal limit) -- \emph{top:} considering the transverse momentum broadening parallel to the magnetic field; \emph{middle:} considering the transverse momentum broadening perpendicular to the magnetic field; \emph{bottom:} sum of the previous contributions for the overall jet quenching of a light parton moving perpendicular to the magnetic field.}
\label{fig3}
\end{figure}

The results for the anisotropic jet quenching parameters in the magnetic EMD model normalized by the conformal limit \eqref{eq:CFT} are presented in Figs. \ref{fig2} and \ref{fig3}.

One concludes that for the magnetic EMD model all the jet quenching parameters increase with increasing values of the magnetic field and, furthermore,
\begin{align}
\hat{q}_{\perp(\parallel)}+\hat{q}_{\perp(\perp)}&\ge\hat{q}_{\parallel(\perp)}\ge \hat{q}_{(\textrm{iso})}\ge\hat{q}_{(\textrm{CFT})},\label{eq:hierarchyEMD1}\\
\hat{q}_{\perp(\perp)}&\ge\hat{q}_{\perp(\parallel)},\label{eq:hierarchyEMD2}
\end{align}
with the equalities being saturated in the limit of zero magnetic field, except for the last inequality on the first line above, which is saturated in the conformal limit attained when the temperature is much larger than any other relevant scale of the model (contrary to the magnetic brane model, the isotropic and the conformal limits are not the same in the magnetic EMD model, because of the dynamical symmetry breaking triggered by the dilaton field).

One also notes from Fig. \ref{fig1} that for all values of the magnetic field, $\hat{q}_{\parallel(\perp)}$ peaks in the crossover region as a function of temperature, while from Fig. \ref{fig2} one sees that for $eB\gtrsim 0.2$ GeV$^2$, $\hat{q}_{\perp(\parallel)}$ and $\hat{q}_{\perp(\perp)}$ are monotonically decreasing functions of temperature.

\section{Conclusions}
\label{conclusion}

In this manuscript I calculated the anisotropic jet quenching parameters for two quite different holographic models at finite temperature and magnetic field. The magnetic brane model, although nonconformal, has no dynamical symmetry breaking, since conformal symmetry is explicitly broken just by the presence of the external magnetic field. Consequently, its phase diagram is a function of the dimensionless combination $eB/T^2$, instead of a function $eB$ and $T$ independently. Although of no direct phenomenological relevance, this is an interesting holographic model since it stems from string theory and corresponds to a strongly coupled anisotropic medium in the presence of a magnetic field. On the other hand, the bottom-up magnetic EMD model does feature dynamical symmetry breaking and its phase diagram is a function of $eB$ and $T$. The magnetic EMD model is a holographic construction of phenomenological relevance for ultrarelativistic peripheral heavy ion collisions, since it is able to correctly \emph{predict} in a quantitative way the behavior of several observables of the strongly coupled anisotropic magnetized QGP, as inferred by comparison with first principles lattice QCD calculations.

Even though these two holographic models at finite magnetic field are fairly different from each other, I found the same general conclusions for the anisotropic jet quenching parameters in both setups. First, there is an overall enhancement of all the jet quenching parameters for light partons with increasing values of the magnetic field. Second, the transverse momentum broadening is larger in transverse directions than in the direction of the external magnetic field. Interestingly, these two conclusions obtained here for light partons are in consonance with the conclusions obtained in Ref. \cite{Finazzo:2016mhm} for heavy quarks, where it was found that the heavy quark energy loss and the Langevin momentum diffusion of heavy quarks are enhanced by the magnetic field, being also larger in transverse directions to the magnetic field. Therefore, these are suggested here as fairly robust features of strongly coupled anisotropic magnetized plasmas.

The results obtained here for the jet quenching parameters of the magnetic EMD model can be employed as microscopic inputs in phenomenological codes for jet quenching and energy loss in the QGP under influence of external magnetic fields.

It would be important to generalize the calculations pursued here for the jet quenching in out-of-equilibrium media, for both models considered. This requires the use of numerical relativity techniques in asymptotically AdS geometries \cite{Chesler:2013lia}. Regarding the magnetic brane model, the homomogeneous isotropization dynamics of this medium has been considered in Ref. \cite{Fuini:2015hba} and it would be interesting to generalize the calculation of anisotropic jet quenching in such setting, and also in inhomogeneous hydrodynamic flows, like the holographic Bjorken flow and shockwave collisions with electromagnetic fields. Ultimately, it would be of great phenomenological interest to carry out such generalizations also for the magnetic EMD model, although that would constitute a much more involved task. I postpone such projects to the future.

\acknowledgments

I acknowledge financial support by Universidade do Estado do Rio de Janeiro (UERJ) and Funda\c{c}\~{a}o Carlos Chagas de Amparo \`{a} Pesquisa do Estado do Rio de Janeiro (FAPERJ).

\end{document}